\documentstyle[12pt]{article}
\def\Ibf{{\bf I}\,}
\def\th{{\theta}}
\def\tpth{\part_{\tb}^p\th}
\def\tqth{\part_{\tb}^q\th}
\def\fpth{\part_{\fb}^p\th}

\def\tpthh{\widehat{\part_{\tb}^p\th}}
\def\tqthh{\widehat{\part_{\tb}^q\th}}
\def\fpthh{\widehat{\part_{\fb}^p\th}}

\def\thh{\hat{\theta}}
\def\thb{\bar{\theta}}

\def\alv{\vec{\alpha} }

\def\tb{\overline{t} }
\def\tbr{\overline{t} }

\def\gam{{ \gamma}\, }
\def\yh{\hat{y}\, }
\def\gamh{\hat{ \gamma}\, }
\def\phih{\hat{ \phi}\, }
\def\mubf{{\bf \mu}\, }

\def\alv{\bf\alpha}

\def\Abf{{\bf A}\,}

\def\Mbf{{\bf M}\,}
\def\Rbf{{\bf R}\,}
\def\Rbfth{{\bf R}_{}^{\theta}\,}
\def\Rbr{{\bf \bar{R}}\,}
\def\muR{{\bf \mu}_{\Rbf}\, }
\def\rharg{{{\bf \rho}({\Rbfth},\mu)} }
\def\muRb{{\bf \mu}_{\Rbr}\, }

\def\fb{\overline{f} }
\def\hb{\overline{h} }

\def\sgh{{\hat \sigma}}

\def\np{\vfill\eject}       
\def\ni{\noindent}
\def\IR{I\kern-.255em R}

\def\lf{\lambda_f }
\def\pb{\overline{p} }

\def\Sbf{{\bf S}\, }

\def\Sbf{{\bf S}\, }
\def\albf{{\bf{\alpha}}\, }

\def\mubf{{\bf{\mu}}\, }
\def\nubf{{\bf{\nu}}\, }
\def\thbf{{\bf{\theta}}\, }

\def\ebf{{\bf e}\,}

\def\rbf{{\bf r}\,}
\def\Abf{{\bf A}\,}

\def\Ibf{{\bf I}\,}

\def\Mbf{{\bf M}\,}
\def\Pbf{{\bf P}\,}

\def\Sbf{{\bf S}\,}

\def\Nth{[{N\over 2}]}
\def\Ab{\bar{A}}
\def\Ah{\hat{A}}
\def\Abf{{\bf{A}}}
\def\Atl{\tilde{A}}
\def\ptw{\tilde{p}}

\def\xtl{\tilde{x}}
\def\ytl{\tilde{y}}
\def\siml{\underline{\sim}}

\def\np{\newpage}
\def\part{\partial}
\def\chtw{\chi_{2}^2}
\def\ynu{y_{\nu}}
\def\sgh{\hat{\sigma}}
\def\yk{y^{(k)}}
\def\Ntw{[{N-1 \over 2}]}
\def\Abr{\bar{A}}
\def\Bbr{\bar{B}}
\def\Dbr{\bar{D}}
\def\Mbr{\bar{M}}

\def\Zbr{\bar{Z}}
\def\ybr{\bar{y}}

\def\Sbf{{\bf{S}}}
\def\Sbr{\bar{S}}
\def\Sh{\hat{S}}
\def\Shb{\bar{\hat{S}}}
\def\Shk{\hat{S}^{(k)}}
\def\Shnu{\hat{S}_{\nu}}

\def\chtw{\chi_{2}^2}
\def\ynu{y_{\nu}}
\def\sgh{\hat{\sigma}}
\def\yk{y^{(k)}}
\def\Ntw{[{N \over 2}]}
\def\Ab{{\bf{A}}}
\def\Eb{{\bf{E}}}

\def\ybr{\bar{y}}

\def\Sbf{{\bf{S}}}
\def\Sbr{\bar{S}}
\def\Sh{\hat{S}}
\def\Shb{\bar{\hat{S}}}
\def\Shk{\hat{S}^{(k)}}
\def\Shnu{\hat{S}_{\nu}}

\begin{document}
\begin{center}
{\bf Optimal Data-based Kernel  Estimation \\
of Evolutionary Spectra}
\end{center}
\begin{center}
{Kurt S. Riedel \\ 
Courant Institute of Mathematical Sciences \\
New York University \\
New York, New York 10012 }\\
\end{center}

\begin{abstract}
Complex demodulation of evolutionary spectra 
is formulated as a two-dimensional kernel smoother in the time-frequency
domain. In the first stage, a tapered Fourier transform,
$y_{\nu}(f,t)$, is calculated. Second, the  log-spectral estimate,
$\thh_{\nu}(f,t)\equiv ln(|y_{\nu}(f,t)|^2)$, is smoothed.  
As the characteristic widths of the kernel smoother increase,
the bias from temporal and frequency averaging increases while
the variance decreases. The demodulation parameters, such as
the order, length, and bandwidth of spectral taper and the kernel smoother,
are determined by minimizing the expected error. 
For well-resolved evolutionary spectra, the optimal taper length is a
small fraction of the optimal kernel halfwidth.
The optimal frequency bandwidth, $w$, for the spectral window scales as
$w^2 \sim \lambda_F/\tau $, where $\tau$ is the characteristic
time, and $\lambda_F$ is the characteristic frequency scalelength.
In contrast, the optimal halfwidths for the second stage kernel smoother
scales as $h \sim 1/(\tau \lambda_F)^{1\over p+2 }$, where $p$ is
the order of the kernel smoother.
The ratio of the optimal frequency  halfwidth to the  optimal
time halfwidth satisfies
${h_F \over h_T} \sim
\left(|\part_{\tb}^p\th| / |\part_{\fb}^p\th| \right)$.
Since the expected loss depends on the unknown evolutionary spectra,
we initially estimate $|\part_{\tb}^p\th|^2$ and $|\part_{\fb}^p\th|^2$
using a higher order kernel smoothers, 
and then substitute the estimated  derivatives into the
expected loss criteria.
\end{abstract}
\newpage
\noindent
{\bf I. Introduction}

In this article, we examine  nonparametric estimates of evolutionary
stochastic processes. We formulate complex demodulation as a
two-dimensional kernel smoother in the time-frequency domain.
In this and a related article, we consider kernel smoother estimates
of time-frequency representations of ``slowly'' evolving time series.
We distinguish three general classes  of time-frequency phenomena.
First, slowly evolving frequency representations are used  when
a quasicoherent signal is most simply represented  in the frequency
domain for intermediate timescales, and the representation evolves temporally.
Second, slowly evolving time representations are used when the signal is most
simply represented as a sum of a small number of sinusoids with time varying
amplitudes. Third, evolutionary spectra are used when the signal is 
incoherent, and can be modeled as a process which is locally stationary
with a slowly evolving spectrum. 

Slowly evolving frequency representations have been described in a number
of excellent review articles (Cohen (1989), Boashash (1990),
Hlawatsch \& Boudreaux-Bartels (1992)),
and will not be investigated in this article.
In a separate article (Riedel (1992b)),  
we considered the problem of estimating
time dependent, deterministic signals in a stationary stochastic
background. In the previous case, only one-dimensional functions
of time or frequency needed to be determined. In contrast,
the amplitude, $A(f,t)$, of an evolutionary process depends on both time
and frequency. Thus the estimation problem is two-dimensional.

Complex demodulation estimates the evolutionary spectral density using a
two step procedure 
(Priestley (1965, 1966), Loynes (1968), Melard \& Herteleer (1989)). 
First, $|A(f,t)|^2$ is estimated on a discrete grid
in time-frequency space using a moving Fourier transform . 
Then a kernel smoother is used to reduce the variance 
of estimate by averaging over a region in the time-frequency domain.
We utilize a {\it two}-dimensional kernel smoother while the standard version
of complex demodulation smoothes only in time. 

Previous studies 
(Priestley (1965, 1966), Papanicolaou et al. (1990), Asch et al. (1991),
Zurbenko (1991))
of complex demodulation emphasized conditions under which the 
evolutionary spectral estimates were asymptotically consistent, i.e.
converged to a consistent estimate as the time scale separation and the
sampling size and rate tend to infinity.
Our study extends these results by determining the leading order bias and
variance of the demodulation.
As the characteristic widths of the kernel smoother increase,
the bias from temporal and frequency averaging increases while
the variance decreases. 

In complex demodulation, both the taper free parameters: 
taper length, taper order, and bandwidth; and the kernel smoother
free parameters: the order, length, and characteristic width of 
the kernel  in each of the time and frequency variables; 
need to be determined. We determine the
demodulation parameters by minimizing the expected error. 
Since the expected loss depends on the unknown evolutionary spectra,
the evolutionary spectra are estimated, and then substituted into the
expected loss criteria. This substitution is known as a ``plug-in''
estimate.

In the next section, we estimate evolutionary spectrum on a 
time-frequency lattice using a moving window Fourier transform and
estimate the bias and covariance of these point estimates. In Section III,
we review kernel smoothers, and determine the bias and variance of the
kernel smoother estimates.
In Section IV, we give asymptotic expressions for the integrated expected loss
and the optimal values of the kernel smoother bandwidths.
In Section V, we describe a class of multiple stage kernel estimators,
which estimate the optimal bandwidth with a pilot kernel estimate.
In Section VI, we briefly discuss the alternatives to two dimensional
kernel smoother estimates of the evolutionary spectrum.  
In the appendix, we examine smoothing kernel estimates
of the coherence and phase of two evolving time series.
An abbreviated version of this article appeared in Riedel (1992a)


\np
\ni
{\bf II. Point Estimates of Evolutionary Processes}

We consider evolutionary processes 
which have the representation:
$$x_t = \int_{-1/2}^{1/2} A(f,t)e^{2\pi ift}dZ(f)
\ \ , \eqno (2.1)$$
where $dZ(f)$ is a weakly stationary stochastic process with 
independent spectral increments and unit variance.
For a given stochastic process, there may be a large family of different
amplitude functions, $A(f,t)$, which satisfy the representation of Eq. 2.1. 
Since $x_t$ is real, we require $dZ(-f) =d\Zbr(f)$ and $A(-f,t)= \Abr(f,t)$.
The evolutionary spectrum is defined as $S(f,t)= |A(f,t)|^2$.
The values of
$dZ$ at different frequencies, $f$ and $f'$, are uncorrelated:
$$
E[dZ(f) d\overline{Z} (f^{\prime} )] =  \delta (f-f^{\prime} )
dfdf^{\prime} 
\ \ .\eqno (2.2)$$
Equation 2.2  implies that the spectral measure is absolutely continuous 
with no deterministic spectral lines.
The unit variance is simply a normalization, 
i.e. we have absorbed the  spectral density, $S(f)$, into $A(f,t)$. 
For stationary processes, $A(f,t)$ is independent of $t$,
and $S(f)=|A(f,t)|^2$. The covariance, $R(t,s)$, satisfies  
$$R(t,s) \equiv {\rm{Cov}}[x_t,x_s] =\int_{-1/2}^{1/2} A(f,t)\Abr(f,s) 
e^{2 \pi i(t-s)f} df 
\ \ .\ \eqno (2.3)$$

We restrict our consideration to oscillatory, semistationary processes
as defined in Priestley (1965). In this context, oscillatory denotes that
$\Ah(f,\hat{f})$ has its maximum at $\hat{f}=0$, where $\Ah(f,\hat{f})$
is the Fourier transform of $A(f,t)$ with respect to $t$ at fixed $f$.
This oscillatory requirement enables us to interpret $A(f,t)$ as the 
time-dependent amplitude for the frequency $f$. Semistationary means that 
$\Ah(f,\hat{f})$ has an uniformly bounded halfwidth,
$\int |\hat{f}||\Ah(f,\hat{f})|d\hat{f}$. Thus the amplitude of a
semistationary process evolves on a bounded timescale.

Even with these restrictions, the amplitude and spectrum may 
not be uniquely defined. 
We also assume $A(f,t)$ is a smooth
function in $C^{\bar{p}}$, where $\bar{p} \geq 2$.
For higher order convergence results, we need to assume $A(f,t)$ is real.

We denote the characteristic time scale of the amplitude
evolution by $\tau$, where we 
normalize the time interval between measurements, $\tau_s$, to unity.  
We make the stronger assumption that
$A(f,t)$ evolves slowly in time and frequency relative to the
sampling rate.
We denote the characteristic frequency scalelength of the
evolutionary spectrum by $\lf$, where 
$\lf^{-2} \siml \part_f^2 S(f,t)/S(f,t)$.
In our asymptotic analysis, we will rescale time and frequency to
the slow scale: $\tb \equiv t /\tau$ and $\fb \equiv f /\lambda_F$.

$2 \tau \lambda_F$ is the number of measurements taken on the timescale
of the spectral evolution divided by the number of degrees of freedom
which are necessary to adequately represent the spectrum.
{\it $ \tau \lambda_F$ is the fundamental expansion parameter}
for complex demodulation. Both the optimal spectral window bandwidth
and the optimal kernel halfwidth are expressed in terms of
$\tau \lambda_F$.

The measured time series is $N_D$ discrete measurements,
$\{ x_0,x_1 , \ldots , x_{N_D-1} \}$,
of a realization of the stochastic process. 
The Nyquist frequency is $1/2$,
and the Raleigh resolution frequency is $1/N_D$.

Our goal is to estimate the local spectral density, 
$S(f,t)\equiv |A(f,t)|^2$, given the measure time series, $\{x_j\}$.
In complex demodulation, we begin the estimate of $|A(f,t)|^2$,
by taking a windowed Fourier transform. For a given taper length, $N$,
and taper, $\nu$,
{\footnote{ In this article, we denote tapers by $\nu$ and kernels by $\mu$.
We normalize tapers to $\sum_j \nu_j^2 = 1$, and  
normalize kernels of zeroth order to $\sum_j \mu_j = 1$.}  
we define the windowed transform, $\ynu(f,t)$ by
$$
y_{\nu}(f,t) = \sum_{j=\Ntw}^{\Ntw} x_{t+j} \nu_j e^{-2\pi if(t+j)} 
\ .\eqno (2.4)$$
For convenience, we assume that the taper length is an odd integer, and
define $\Ntw \equiv ({N-1 \over 2})$.
The point estimate of the evolutionary spectrum is
$\Shnu(f,t) = |\yh_{\nu}(f,t)|^2$. 
Equation 2.1 implies that $\ynu(f,t)$ satisfies the statistical model
$${y}_{\nu}(f,t) = \int_{-1/2}^{1/2}\sum_{j=\Ntw}^{\Ntw} 
A(f',t+j) \nu_j e^{2\pi i(f'-f)(t+j)}dZ(f') 
\ .\eqno (2.5) $$
Since we assume that $A(f,t)$ is a smooth, slowly varying function,
we make a Taylor series expansion of $A(f,t+j)$  about $(f,t)$:
$A(f,t+j)= A(f,t) +\part_t A(f,t)j + O(|A|j^2/\tau^2)$.
With this expansion, Eq. 2.5 reduces to
$$
{y}_{\nu}(f,t) = \int_{-1/2}^{1/2}\sum_{j=\Ntw}^{\Ntw} 
[A(f',t)+\part_t A(f',t)j] \nu_j e^{2\pi i(f'-f)(t+j)}dZ(f') 
\ .\eqno (2.6) $$
To simplify Eq. 2.6,
we define the spectral window, $V(f)$, to be the Fourier transform of $\nu_j$:
$$
V(f) = \sum_{j=-\Ntw}^{\Ntw} \nu_j e^{-2 \pi ijf} \ ,
\ \ {\nu}_j = \int_{-1/2}^{1/2} V(f) e^{2 \pi ijf}df . \eqno (2.7)
$$
The uniform taper, $\nu_j \equiv{ 1\over \sqrt{N}}$, 
generates the spectral window,
$V(f) = {1\over\sqrt{N}}\sum_{j=-\Ntw}^{\Ntw}  e^{-2 \pi ijf}$ 
$= {1\over\sqrt{N}}e^{ \pi i(N-1)f}D_N(f)$,
where $D_N$ is the Diriclet kernel. In the frequency domain, 
Eq. 2.6 becomes a convolution equation:
$${y}_{\nu}(f,t) = \int_{-1/2}^{1/2}
[A(f',t)V(f-f')-{ \part_t A(f',t)\part_fV(f-f')\over 2\pi i}] 
\ e^{2\pi i(f'-f)t}dZ(f') 
\ .\eqno (2.8) $$
Thus the discrete Fourier transform of the measured process is related
to the realization of the stochastic process by an integral equation of the
first kind. 
The purpose of the taper is to reduce the bias from the sidelobes 
of the kernel, $V(f-f')$, relative to those of the Diriclet kernel.
We denote the characteristic bandwidth of the spectral kernel by $w$. 
Typically, the taper is chosen with $w \siml c/N\tau_s$, where $c \ge 1$.  

The expectation of the quadratic tapered estimator is
$${\bf E} \left[  ({y}_{\nu}(f_1,t){\ybr}_{\nu}(f_2,t') \right]
= e^{2\pi i(f_2t'-f_1t)}\int_{-1/2}^{1/2} 
\left[ M(f',t,f_1)\Mbr(f',t',f_2)e^{2\pi if'(t-t')}
\right] df' \ , \eqno (2.9)$$ 
where $M(f',t,f) \equiv A(f',t)V(f-f')  \ - 
{ \part_t A(f',t)\part_fV(f-f')\over 2\pi i}$.
Thus ${y}_{\nu}(f_1,t)$ and ${\ybr}_{\nu}(f_2,t')$
become nearly independent when $|f_1 -f_2|> w$ or  $|t-t'| >{1\over w}$,
where $w$ is the spectral bandwidth.
When $f_1=f_2$ and $t = t'$, Eq. 2.9 reduces to
$${\bf E} \left[  ({y}_{\nu}(f_1,t){\ybr}_{\nu}(f_1,t)) \right]
=
\int_{-1/2}^{1/2} 
\left[ |A(f',t)|^2|V(f'-f_1)|^2 \ +
{|\part_tA(f',t)|^2|\part_fV(f'-f_1)|^2 \over 4\pi^2 }\right] df' 
\ . \eqno (2.10)$$ 
The second term in Eq. 2.10 constitutes the bias in the spectral estimate
due to the temporal variation in the amplitude $A(f,t)$.
This bias is always positive. The bias  
from the frequency variation in $S(f,t)$ is estimated by
making a Taylor series expansion of $S(f,t)$.
The total bias of the tapered spectral estimate, $\Shnu(f,t) \equiv
|{y}_{\nu}(f,t)|^2$ is approximately 
$${\rm Bias} \left[\Shnu(f,t)\right] \siml 
\part_{\fb}^2 S(f,t) \Bbr_{\nu} ({w\over \lambda_F})^2 +
|\part_{\tbr}A(f',t)|^2 \Dbr_{\nu} ({1 \over \tau w})^2
\ , \eqno (2.11)$$
where
$$ \Bbr_{\nu} \equiv {1 \over w^2}\int_{-1/2}^{1/2} |f|^2 |V(f)|^2 df
\ , {\rm and} \
\Dbr_{\nu} \equiv {w^2 \over 4\pi^2} \int_{-1/2}^{1/2} |\part_f V(f)|^2 df
\ . \eqno (2.12)$$
We note that ${\rm{B}} [\thh_{\nu}(f,t)]$ contains a term which is proportional
to $|\part_t A(f,t)|^2$. We estimate only $S(f,t)$,
so $|\part_t A(f,t)|^2$ is undetermined unless we assume that
$|\part_t A(f,t)|^2 \sim |\part_t S(f,t)|^2/ 4S(f,t)$.
{\it The bias of the point estimate of $S(f,t)$ is minimized by
choosing the taper bandwidth to minimize Eq. 2.11, i.e.
$w^2 \sim \lambda_F/\tau $} and the taper length, $N$, satisfies
$N^2  \sim {\tau \over \lambda_F\tau_s^2}$.}
Due to symmetry, this expression is valid to
to fourth order in the small parameters, $w\over \lambda_F$
and $1\over \tau w$.


Since $|y_{\nu}(f,t)|^2$  is a narrow bandpass process, it 
is approximately distributed as a $\chi_2^2$ distribution 
for a wide class of distributions, $dZ(f)$ (Mallows (1967)). 
When $dZ(f)$ are {\it Gaussian} variables, we
estimate the covariance of $\Sh(f,t)$  by applying
Isserlis'  fourth moment identity for complex { Gaussian} variables:
$$ E\left[ X_1X_2X_3X_4\right] =
E\left[ X_1X_2\right]  E\left[ X_3X_4\right] +
E\left[ X_1X_3\right]  E\left[ X_2X_4\right] +
E\left[ X_2X_3\right]  E\left[ X_1X_4\right] .$$
For incremental Gaussian processes, $dZ(f)$, we have
$$ {\bf Cov} \  [|{y}_{\nu}(f_1,t)|^2, |{y}_{\nu}(f_2,t')|^2] =
\Eb[|{y}_{\nu}(f_1,t)|^2 |{y}_{\nu}(f_2,t')|^2] -
\Eb[|{y}_{\nu}(f_1,t)|^2 ] \Eb[|{y}_{\nu}(f_2,t')|^2] = 
$$ $$
\left|{\bf E} \  [{y}_{\nu}(f_1,t){\ybr}_{\nu}(f_2,t')]\right|^2 +
\left| {\bf E} \  [{y}_{\nu}(f_1,t'){\ybr}_{\nu}(-f_2,t')]\right|^2 
\ , \eqno (2.13)$$
where ${\bf E} \  [{y}_{\nu}(f_1,t){\ybr}_{\nu}(f_2,t')]$ is given
by Eq. 2.9.
The second term  is only important when 
both $f_1$ and $f_2$ are within a bandwidth of zero frequency. 
Thus the variance of the estimated spectrum is
$$
{\bf Var} \  [\Sh_{\nu}(f_1,t)]  \siml ( 1 +\delta_{f,0})
 \left| S(f,t) + \part_{\fb}^2 S(f,t) \Bbr_{\nu} ({w\over \lambda_F})^2 +
|\part_{\tb}A(f',t)|^2 \Dbr_{\nu} ({1 \over \tau w})^2
\right|^2 
\ .\eqno (2.14)
$$
Thus $\Sh_{\nu}(f,t)$  has a variance almost exactly
equal to the square of its expectation away from $f=0$.   

Although the windowed transform of Eq. 2.4 is defined on all points
$(f,t),\ |f|\le 1/2$, we only evaluate $\ynu(f,t)$ on a discrete lattice,
$t= j \delta t,\ f=m \delta f$, where $j = 0,\ldots N_t, \
m=0,N_f$, where $N_t \equiv N_D/ \delta t$, $\delta t \equiv N p_t\tau_s$, 
$N_f \equiv N / p_f$ and $\delta f \equiv  {p_f\over N\tau_s}$. 
This time-frequency lattice is precisely the Gabor transform of the
discrete process.
$(1 - p_t)$ and $(1 - p_f)$ are the grid overlap fractions for time and
frequency. Equation 2.7 shows that the lattice point estimates become
nearly independent as $p_t$ or $p_f$ approach one. 
As the overlap fraction decreases, the additional lattice points contain
less new information and more redundant information. 

\newpage
{\bf III. Bias and Variance of Smoothed Kernel Estimates}

The point estimate for the spectral density, $\Sh(f,t) = |\ynu(f,t)|^2$,
is inconsistent, i.e. the variance of the estimate  does 
not tend to zero as the number of datapoints, $N_D$, increases.
Statistical consistency is normally achieved 
by smoothing the point spectral estimate over a region.
Smoothing estimators may broadly be 
distinguished by three main characteristics:
dependent variable, smoothing method, and data-based parameter selection
technique. Prior to smoothing, it is usually desirable to transform 
the spectrum to the logarithmic scale: $\thh(f,t) \equiv ln(\Sh(f,t))$,
and then transform back both the smoothed estimate and the confidence interval.
The logarithmic transformation is advantageous, because $\thh$ has approximately
a $log(\chi_2^2)$ distribution, which is closer to a Gaussian than $\chtw$ is.
Also the variance of $\thh$ is independent of $\th$, while the variance of
$\Sh(f,t)$ is equal to $(E[\Sh(f,t)])^2$.  When $\yh$ has a Gaussian 
distribution, $\thh(f,t)$ has expectation and variance:
$${\rm E}[\thh(f,t)] =\ \th(f,t) + \psi(1) \siml\ \th(f,t) -.5772 \  , \  \
{\rm Var}[\thh(f,t)] = \psi'(1) \siml\ 1.645
\ ,\eqno (3.0)$$
\ni
where $\psi$ is the digamma function.
More detailed discussions of variance stabilizing transformations can be
found  
in Thomson and Chave (1990). We bias correct  the tapered log-periodogram
prior to smoothing: $\thh_{new}(f,t) = \thh_{raw}(f,t) + .5772$.

The standard smoothing methods are kernel smoothers and smoothing splines.
We concentrate on kernel smoothers because both the error analysis and
the implementation are easier. In Section  VI,
we discuss the smoothing spline alternative.

A data-based method to determine the smoothing free parameters
needs to be selected. In Wahba (1980), generalized cross-validation,
a statistical resampling technique, is advocated to determine the
level of smoothing. However,
data-based parameter selection methods which minimize an estimate of
the expected loss converge more rapidly than
generalized cross-validation. 
Thus we will estimate the bias and variance of the kernel smoother,
and then select the smoother parameters to minimize this estimate
of the expected loss.

We consider two-dimensional kernel smoothers which are the
crossproduct of a kernel smoother in frequency, 
$\mu_{F}$, and a kernel smoother in time, 
$\mu_{T}$. Before examining two-dimensional kernels, we review 
the basic definitions and properties of one-dimensional kernels.
We index the one-dimensional kernel by the index $j$, where $|j| \le M$.
We call $M$ the index bound.
We say that a one-dimensional kernel, $\mubf$, has halfwidth, $H$, 
and is of type $(q,p)$ if
$$
\mubf \cdot\rbf_{m} =  { q! H^q}\delta_{m,q} \ , \ m= 0, \ldots,p-1
,\eqno(3.1)$$
where
$\rbf_{m,j} \equiv j^{m} , \  -M\le j \le M$.
We denote the $p$th moment by $\mubf \cdot\rbf_{p} =   p! C(q,p)H^p$,
and the $(p+\ptw)$th moment by 
$\mubf \cdot\rbf_{p+\ptw} = (p+\ptw)! C_{\ptw}(q,p)H^{p+\ptw}$,
where $\ptw$ is a positive integer.
Kernels of type $(q,p)$ are used to estimate the $q$th derivative
of a function, $g({j \delta t \over \tau})$, to order 
$O(({H\tau_s\over\tau  })^{p-q})$.
Typically, the index bound will be a multiple of the halfwidth.
Most widespread kernels are scale parameter kernels, which have 
the form $\mubf_j = {1 \over H}K({j \over H})$.

We assume that the crossproduct kernel smoother has the same order $p$
in time and in frequency.
We let $H_T$ and $H_F$ denote the halfwidths, and $M_T$ and $M_F$ denote 
the index bounds of the kernel smoothers, $\mu_{T}$ and $\mu_{F}$.
We also define the normalized halfwidths, $h_F$ and $h_T$, by
$h_T ={H_T \delta t\over \tau}= {H_T p_t N\tau_s\over \tau}$ and 
$h_F ={H_F \delta f\over \lambda_F}= {H_F p_F \over \lambda_F N\tau_s}$.
If $\mu_T$ has type $(q,p)$
and $\mu_F$ has type $(q',p')$, the corresponding results can be 
derived by replacing all occurences of $h_F^{p}$ and $h_F^{q}$
by $h_F^{p'}$ and $h_F^{q'}$.

The two-dimensional crossproduct kernel estimators have the form:
$$
\hat{\theta}_{\mu}(f,t) =
\sum_{k=-M_F}^{M_F} \sum_{j=-M_T}^{M_T}
\mubf_{F,k}\mubf_{T,j} \thh_{\nu}(f+k\delta f,t+j\delta t)
\ .\eqno (3.2) $$
We also need to estimate time and frequency derivatives of $\th(f,t)$.
To estimate the $q$th time derivative of $\th(f,t)$, we use a
crossproduct kernel where $\mu_T$ is of type $(q,p)$
and $\mu_F$ is of type $(0,p)$.
We denote the estimate of $\partial_t^{q}\theta(f,t)$
by $\widehat{\partial_t^{q}{\theta}}(f,t)$.

We divide the expectation of the point estimate,
$\thh_{\nu}(f,t)$, into $\th(f,t)$ and the bias, $B[\thh_{\nu}(f,t)]$:
$E[\thh_{\nu}(f,t)] = \th(f,t)+ B[\thh_{\nu}(f,t)]$,
where $B[\thh_{\nu}(f,t)]$ 
$\siml {d \th \over dS} ({\thh_{\nu}}(f,t))B[\Sh_{\nu}(f,t)]$
$\siml [ \part_{\fb}^2 \th(f,t) + (\part_{\fb} \th(f,t))^2] \Bbr_{\nu}
({w\over \lambda_F})^2 + |\part_{\tb}\th(f,t)|^2 
\Dbr_{\nu}({1\over 2\tau w})^2$.
We order the bias of the point estimate to be smaller than the bias
of the kernel smoother. We can then approximate the bias of
$\thh_{\mu}(f,t)$ to high order accuracy by:
$$B[\thh_{\mu}(f,t)] \siml 
C(0,p) \part_{\fb}^p \th(f,t)h_F^p + C(0,p) \part_{\tb}^{p} \th(f,t)h_T^p +
B[\thh_{\nu}(f,t)] +
$$ $$ 
C_2(0,p) \part_{\fb}^{p+2} \th(f,t)h_F^{p+2}
+ C_2(0,p) \part_{\tb}^{p+2} \th(f,t)h_T^{p+2} +
C(0,p)^2 \part_{\fb}^p \part_{\tb}^{p} \th(f,t)h_F^{p}h_F^{p}
\ .\eqno (3.3)$$ 
The last term is important only when $p$ equals two.

To estimate the $q$th time derivative of $\th(f,t)$, we use a
crossproduct of a kernel smoother in time of type $(q,p)$,
and a kernel smoother in frequency of type $(0,p)$. 
For a kernel smoother 
of type $(q,p)$, the bias of the estimate of $\widehat{\part_t^q\th(f,t)}$ is
$$B[\widehat{\part_{\tb}^q\th_{\mu}}(f,t)] \siml 
C(0,p) \part_{\fb}^p \th(f,t){h_F^p\over h_T^q} 
+ C(q,p) \part_{\tb}^{p} \th(f,t)h_T^{p-q} +
{\part_{\tb}^qB[\thh_{\nu}(f,t)]} \  +
$$ $$ 
C_2(0,p) \part_{\fb}^{p+2} \th(f,t){h_F^{p+2}\over h_T^q}
+ C_2(q,p) \part_{\tb}^{p+2} \th(f,t)h_T^{p+2-q} +
C(q,p)C(0,p) \part_{\fb}^p \part_{\tb}^{p} \th(f,t)h_F^{p}h_T^{p-q}
\ ,\eqno (3.4)$$ 
which includes Eq. 3.3 as the special case $q=0$.
Reversing the time and frequency indices yields the corresponding
results for frequency derivatives.

We let $R_{\nu}^{\th}(f_1,t_1,f_2,t_2)$ denote the covariance of the
transformed spectral density: 
$$R_{\nu}^{\th}(f_1,t_1,f_2,t_2)\equiv 
{\rm Cov}[{\thh_{\nu}}(f_1,t_1),\thh_{\nu}(f_2,t_2)]
\ .\eqno (3.5)$$
Eqs. 2.9 \& 2.13 give the covariance of 
$\Sh_{\nu}(f_1,t_1),\Sh_{\nu}(f_2,t_2)$ in terms of $A(f,t)$.
When $A(f,t)$ is unknown, the covariance of 
${\th_{\nu}}(f_1,t_1)$ and $\th_{\nu}(f_2,t_2)$ may be estimated with 
the plug-in approximation. 

We would like to be able to replace
${\rm Cov}[{\thh_{\nu}}(f_1,t_1),\thh_{\nu}(f_2,t_2)]$
by its Taylor series approximation:
$${\rm Cov}[{\thh_{\nu}}(f_1,t_1),\thh_{\nu}(f_2,t_2)] \siml
C {d \th \over dS} ({\thh_{\nu}}(f_1,t_1))
{d \th \over dS} (\thh_{\nu}(f_2,t_2))
{\rm Cov}[{\Sh_{\nu}}(f_1,t_1),\Sh_{\nu}(f_2,t_2)]
\ . \eqno (3.6)$$
We choose the constant $C$ in Eq. 3.6 to match known results at
$(f_2,t_2) = (f_1,t_1)$. 
Although Eq. 3.6 is not strictly valid, this simplification is widespread 
due to the severe complications of calculating 
${\rm Cov}[{\thh_{\nu}}(f_1,t_1),\thh_{\nu}(f_2,t_2)]$ exactly.
If the multiple taper spectral estimates of Thomson (1982, 1990) 
are used, 
the expansion parameter scales as the inverse of the number of tapers.

To further simplify the evaluation of
${\rm Cov}[{\thh_{\nu}}(f_1,t_1),\thh_{\nu}(f_2,t_2)]$, we expand
$M(f',t_{\ell},f_{\ell})$ in Eq. 2.9 by
$A(f_{\ell},t_{\ell})V(f_{\ell}-f')  \ - 
{ \part_t A(f_{\ell},t_{\ell})\part_fV(f_{\ell}-f')\over 2\pi i}$,
where the subscript ${\ell}$ equals one or two. 
Thus for the $log$ transformation, we have
$${\rm Cov}[{\thh_{\nu}}(f_1,t_1),\thh_{\nu}(f_2,t_2)] \siml
C\ \left| \int_{-1/2}^{1/2} \left[ V(f_1-f')V(f_2-f')
e^{2\pi if'(t-t')}\right] df' \right|^2
\ , \eqno (3.7)$$
to the second order in the small parameters, $w\over \lambda_F$
and $1\over \tau w$. From Eq. 3.0, we choose $C =\psi'(1) \siml\ 1.645$. 
When the grid overlap parameters, $p_T$ and $p_F$, are greater than one,
${\rm Cov}[{\thh_{\nu}}(f_1,t_1),\thh_{\nu}(f_2,t_2)]$ is approximately
equal to a multiple of the identity
matrix. When $p_T$ and $p_F$ are less than one,
the coupling matrix needs to be calculated.

Given the covariance of the point estimate, $R_{\nu}^{\th}(f_1,t_1,f_2,t_2)$,
the variance of the smoothed kernel estimator 
of Eq. 3.2 is
$$ {\rm Var} \ [\widehat{\part_{\tb}^q\th_{\mu}}(f,t)] = {1 \over h_T^{2q}}
\sum_{j,j'=-M_T}^{M_T}\sum_{k,k'=-M_F}^{M_F} 
\mubf^T_j \mubf^F_k R_{\nu}^{\th}(f+k,t+j,f+k',t+j') \mubf^T_{j'}\mubf^F_{k'} 
\  \ .\eqno (3.8)$$
{\it The righthand side of Eq. 3.6 scales as 
${ (\tau \lambda_Fh_F h_T^{2q+1}})^{-1}$ 
times the variance of $\thh_{\nu}(f,t)$.}
To simplify later formulas, we  denote the righthand side of Eq. 3.8 by 
${\rharg / (h_F h_T^{2q+1})}$. 
$\tau \lambda_F$ (or more precisely, $(\tau \lambda_F)^{1\over 2p+2}$
is the ratio of the sampling rate to
the number of degrees of freedom
which are necessary to adequately represent the spectral evolution.
As such, $ \tau \lambda_F$ is the fundamental expansion parameter
for complex demodulation.

\np
\noindent
{\bf IV. Optimization of Expected Loss for Complex Demodulation}

The local expected loss of the smoothed estimate of the $q$th derivative
is the sum of the squared bias and the variance:
$$
L(\widehat{\part_{\tb}^q\th_{\mu}}(f,t), \mubf ) = 
|B[\widehat{\part_{\tb}^q\th_{\mu}}(f,t)]|^2
+ {\rharg \over h_F h_T^{(2q+1)} }
\ , \eqno(4.1)$$
where the bias is given by Eq. (3.4) and $\rharg = O({1\over \tau\lambda_F})$.
Equation 4.1 shows that
as the characteristic widths of the kernel smoother increase,
the bias from temporal and frequency averaging increases while
the variance decreases. 
The bias versus variance tradeoff is well known in smoothing
kernel estimates of stationary spectra
(Grenander and Rosenblatt (1957), Wahba (1980)).
For evolutionary spectra, a similar expression for kernel smoothing only in
time is given by Priestley (1966).

To simplify the expected loss, we expand Eq. 4.1 in powers
of $h_T$ and $h_F$. The leading order expected loss is 
$$
L_W^o(\widehat{\part_{\tb}^q\th_{\mu}}(f,t), \mubf ) \ =
\left[ C(0,p) \part_{\tb}^p\th_{\mu}(f,t){\hb_F^{p} \hb_T^{-q}} +
 C(q,p)\part_{\fb}^p\th_{\mu}(f,t)\hb_T^{p -q} \right]^2 
+ {\rharg \over h_F h_T^{(2q+1)} }
\ . \eqno(4.2)$$
To the leading order, only ${\part_{\tb}^p\th_{\mu}}(f,t)$ and  
${\part_{\fb}^p\th_{\mu}}(f,t)$ appear in the loss criterion, and
the mixed partial derivatives are not required.
To determine the leading order optimal values of the kernel scale parameters,
we minimize Eq. 4.1 with respect to the $h_T$ and $h_F$.
We begin by changing variables to $h\equiv \sqrt{h_T h_F}$
and $r\equiv \sqrt{{h_F \over h_T}}$. The minimizing value of $r$ 
is order one, and satisfies:
$$
\left[ C(0,p) \part_{\tb}^p\th_{\mu}{r^{p}} +
 C(q,p)\part_{\fb}^p\th_{\mu}r^{-p} \right]
\left[(2pq +p+q)C(0,p) \part_{\tb}^p\th_{\mu}{r^{p}} -
(p-q) C(q,p)\part_{\fb}^p\th_{\mu}r^{-p} \right]
\ . \eqno(4.3)$$
The optimal value of $h$, denoted by $h_o$, satisfies 
$$
h_{o}(\mubf )^{2p+2} = {q+1 \over (p-q)}
{\rharg \over K(r)} \ , \ 
{\rm where}\ \ K(r) \equiv \left[ C(0,p) \part_{\tb}^p\th_{\mu}{r^{p}} +
 C(q,p)\part_{\fb}^p\th_{\mu}r^{-p} \right]^2
\ .\eqno(4.4)$$
Equation 4.4 has two roots; when  
$C(0,p) C(q,p)\part_{\tb}^p\th_{\mu}{r^{p}} \part_{\fb}^p\th_{\mu} > 0$,
we have 
$${r} = \left[{ (p-q)\over(2pq +p+q)}
{ C(q,p)\part_{\fb}^p\th_{\mu}r^{-p} \over
C(0,p) \part_{\tb}^p\th_{\mu} 
} \right]^{1\over  2p}\ . \eqno(4.5)$$
When $C(0,p) C(q,p)\part_{\tb}^p\th_{\mu}{r^{p}} \part_{\fb}^p\th_{\mu} < 0$,
the `optimal' halfwidth satisfies $K(r) = 0, \ h = \infty$!
This pathological result occurs because we have neglected the higher order
bias terms, and thus $K(r)= 0$ eliminates our approximate bias.
In some, but not all cases, including the next order terms from Eq. 3.4
will remove the singularity and result in  a  finite value of $h$.
Including these higher order terms increases the computational cost
and algorithmic complexity and often causes ill-conditioning. 
{\it Therefore we regularize the Eq.  4.1-3 by replacing $K(r)$ with a
function, $K_b(r)$, which is bounded from below by a positive
constant.} Remark B gives one possible regularization.   
In the future, we plan to explore the effect of different regularization
schemes on the evolutionary spectral estimate.

In general, the optimal value of $h$ is proportional to 
$(\tau\lambda_F)^{-1/(2p+2)}$,
and the total error, $ L^2(\partial_t^q \thh;\mubf )$,
is proportional to $(\tau\lambda_F)^{-2(p-q)/(2p+2)}$.
If $\th(f,t)$ has precisely $\pb$ continuous derivatives, then kernels of type
$(q,\pb)$ yield the highest possible rate of convergence asymptotically
as $\tau \lambda_F$ increases (Stone (1982)).  
For given time and frequency scalelengths,
 and a given sampling rate,  higher order kernels
may produce worse estimates, because higher order kernels require
larger values of $\tau \lambda_F$ and $N_D$ to be in the
asymptotic limit.
When $C(0,p) C(q,p)\part_{\tb}^p\th_{\mu}{r^{p}} \part_{\fb}^p\th_{\mu} < 0$
and  $K(r)=0$, the kernel estimator is effectively higher order.
By regularizing $K(r)$, we maintain a  kernel of $\bar{p}$ order.

{\it Remarks}:

{\it A) Edge kernels}

When the domain of the kernel smoother intersects the ends of the
dataset, the kernel needs to be  modified to continue to be of type 
$(q,p)$. The appropriate edge kernels are given in Riedel \& Sidorenko (1993).
Although the edge kernels noticably increase the computation cost and 
implementation complexity, they yield  significantly better results 
in practice.
 
{\it B) Regularization of the optimal halfwidths}

A desirable regularization is to integrate Eq. 4.1 over
a small region in $(f,t)$ space. To model this effect,
we replace Eq. 4.2  by
$$ (2pq+p +q)C(0,p)^2 I_{F,F}^{p,p} r^{2p}  +
2q(p+1)C(0,p) C(q,p)I_{T,F}^{p,p} 
- {(p-q)}C(q,p)^2 I_{T,T}^{p,p} r^{-2p} = 0 
\  .\eqno(4.6)$$
where  
$$I_{T,T}^{p} \equiv |\part_{\tb}^p\th(f,t)|^2 \ ,\ 
I_{F,T}^{p} \equiv b \part_{\fb}^p\th(f,t) \part_{\tb}^{p'}\th(f,t)\ ,\ 
I_{F,F}^{p} \equiv |\part_{\fb}^p\th(f,t)|^2 
, \eqno(4.7)$$
with $0< b<1$. If  we were to integrate Eq. 4.1 over a small region,
the regularization parameter, $b$, would  be
$$ b \siml
I_{F,T}^{p,p'}(\th,W) \equiv {
\int df \int dt \part_{\fb}^p\th(f,t) \part_{\tb}^{p'}\th(f,t)\over 
\left[\int df \int dt |\part_{\tb}^p\th(f,t)|^2\right]^{1\over 2}
\left[\int df \int dt |\part_{\fb}^p\th(f,t)|^2\right]^{1\over 2}
}. \eqno(4.8)$$
Equation 4.8 is used only heuristically for scalelength adjustment
and is not actually applied.

{\it C) Global halfwidth kernels}

In this article, we consider  variable, locally adaptive halfwidths.
If instead the halfwidth is fixed globally, the convergence rate
will be suboptimal, and the difference in the expected losses is given
by a Holder inequality . 
The more rapidly $\part_{\tb}^p\th(f,t)$ and $\part_{\fb}^p\th(f,t)$ vary, 
the larger the difference will be (Mueller \& Stadtmueller (1987)).

{\it D) Selection of the grid overlap parameters, $p_T $ and $p_F$}

Decreasing the grid overlap parameters, $p_T $ and $p_F$,
will decrease the expected loss because no information is lost
and a small amount is gained. However, the decrease in the loss
becomes very weak when $p_T$ and $p_F$ are decreased below one third.
Minimizing the loss function with respect to the grid overlap parameters 
will result in the trivial and unachievable minimum at $p_T =0$ and $p_F=0$.
Thus the tradeoff in overlap parameter selection is expected loss
versus computational effort.

\ \

{\bf V. Pilot parameter determination for complex demodulation}

To optimize the final estimate of $\th(f,t)$, we consider multiple stage
estimators. We classify the multiple stage kernel estimators by:
A) the kernel order, $p$, of the  final stage estimate;
B)  the number of stages, where each earlier stage estimates 
the optimal halfwidths
for the next stage;
C) the type of initialization.

The combined scheme is constrained to not estimate derivatives of order
greater than the number of continuous derivatives, $\bar{p}$, of $\th(f,t)$. 
Higher order kernels can yield poor
results when $\th(f,t)$ is less smooth than the order of
the kernel. Therefore, we check $a$ $posteori$ that the estimated 
derivatives have no discontinuities.
For each of these multiple stage estimates, {\it we evaluate the expected
loss using Eq. 4.2 with the plug-in approximation.
We then select the scheme with the smallest expected error.}
Multiple stage kernel estimators with variable kernel halfwidths
are discussed for one dimensional problems in
Mueller \& Stadtmueller (1987) and 
Riedel (1992b).

Given estimates of ${\part_{\tb}^p\th_{\mu}}(f,t)$ and 
${\part_{\fb}^p\th_{\mu}}(f,t)$, 
we  construct an estimate of the optimal
halfwidths, $h_T$ and $h_F$, by substituting the estimates,
$\widehat{\part_{\tb}^p\th}(f,t)$ and 
$\widehat{\part_{\fb}^p\th}(f,t)$,
into a regularized version of Eqs. 4.2-3. 
Provided that ${\part_{\tb}^p\th}(f,t)$ and
${\part_{\fb}^p\th}(f,t)$
are continuous, the empirical
halfwidth estimate of ${\part_{\fb}^q\th}(f,t)$ is accurate to
$(\tau\lambda_F)^{-(p-q)/(2p+2)}$, i.e.
$ L^2(\widehat{\partial_t^q \th}) \siml O(( \tau\lambda_F)^{-2(p-q)/(2p+2)})$.
Thus the rate of convergence is optimal. However,
the constant of proportionality is suboptimal. More precisely, 
 $${\rm E}\left[|\tqthh(\hat{h}_{q,p}) - \tqth|^2 \right]
\sim c(\tau\lambda_F) 
{\rm E}\left[|\tqthh({h}_{q,p}) - \tqth|^2 \right]
\ ,\eqno(5.1)$$
where $h_{q,p}$ is given by Eqs. 4.3-4  and  
$\hat{h}_{q,p}$ is the empirical estimate. If 
$c(\tau\lambda_F)\rightarrow 1$ as $\tau\lambda_F \rightarrow \infty$,
the empirical estimate, $\tqthh(\hat{h}_{q,p})$, of $\tqth$ 
is asymptotically efficient.
The relative convergence rate is the rate which 
$c(\tau\lambda_F)-1$ tends to zero. Since
$ L^2(\widehat{\partial_t^q \theta})$ is a smooth function of $h_T$ and $h_F$,
we have
$$c(\tau\lambda_F) \siml \ 1 \  + O( |{|\tpthh|^2 \over |\tpth|^2} -1|) +
 O(|{|\fpthh|^2 \over |\fpth|^2} -1)
\ .\eqno(5.2)$$
Thus to optimize the relative rate of convergence in our estimate
of $\tqth$, we begin by estimating 
$\tpth$ and $\fpth$ in an earlier stage.

To optimize the final estimate of $\th(f,t)$, we consider multiple stage
estimators. In two stage schemes, we begin by estimating $\tpth(f,t)$
and $\fpth(f,t)$ using crossproduct kernels of type 
$(0,p+2)\times (p,p+2)$ and $(p,p+2)\times (0,p+2)$ respectively.
 The kernel halfwidths for the first
stage are determined by substituting into Eq. 4.2-3 
the {\it scalelength ansatz:}
$\widehat{\part_{\tb}^{p+2}\th}(f,t)\sim \tau^{-p-2}\thb$ 
and $\widehat{\part_{\fb}^{p+2}\th}\sim \lambda_F^{-p-2}\thb$
where $\thb$, $\tau$ and $\lambda_F$ are given $a$ $priori$. 
In the second stage, we estimate $\th(f,t)$ using
a $(0,p)\times (0,p)$ kernel with the halfwidths specified by
$\tpthh(f,t)$ and $\fpthh(f,t)$.

The final two stages of three stage estimators are the kernel estimates of
$\tpth(f,t)$, $\fpth(f,t)$, and $\th(f,t)$ as in two stage estimates.
There are three types of initializations for three stage estimators,
scalelength ansatz estimators, goodness of fit/ factor method estimators,
and parametric estimators.
Each type estimates the optimal halfwidths for the second  stage.
Scale length ansatz estimators  begin by applying 
$(0,p+4)\times (p+2,p+4)$ and $(p+2,p+4)\times (0,p+4)$
kernels with the halfwidths specified by Eqs. 4.2-3 and using
the scalelength ansatz for the $(p+4)$th partial derivatives.
In three and four stage estimators, considerable error can be tolerated
in the initial stages without noticibly impacting the final estimates.

The goodness of fit/factor method (Mueller \& Stadtmueller (1987))
initializes by estimating the optimal global halfwidth for a 
$(0,p+2)\times (0,p+2)$ kernel using the Rice goodness of fit
criterion (Hardle et al. (1988)).
The Rice criterion selects the  halfwidths to  minimize
$$ C_R(h_F,h_T)\equiv
\sgh(h_{0,p'})^2
(1 - {2\mu_{0}\over N_D} ) \ ,  \ \eqno(5.3)$$ 
$${\rm where} \ \ \sgh(h_{0,p'})^2 \equiv 
\sum_{j=0}^{N_t}\sum_{k=0}^{N_f} 
{\left[ln(|y_{\nu}(m\delta f, j \delta t)|^2) -
\thh_{\mu} (m\delta f, j \delta t)  \right]^2 \over N_t N_f}
\ .\eqno(5.4)$$
In Equation 5.3,
$\sgh(h_{0,p'})^2$
is the mean squared residual error and $(1 - {2\mu_{0}\over N_D} )$ 
corrects for the number of degrees of freedom. 
The Rice goodness of fit criterion is an asymptotically efficient
estimator of the optimal halfwidth, but it has a slow relative
rate of convergence (in comparison with the multiple stage kernel estimators). 

After the minimum of Eq. 5.3, $\hat{h}_{0,p'}$, is found, 
the factor method selects halfwidths for derivatives, $h_{q',p'}$,
in terms of the optimal halfwidth for a $(0,p')$ kernel:
$$ h_{q',p'} = H(\mu_{q',p'},\mu_{0,p'}) \hat{h}_{0,p'} , \ \ 
{\rm where} \  H(\mu_{q,p},\mu_{0,p}) \equiv 
\left({ (4pq+2p) C(0,p)^2 m_{2}(\mu_{q,p})
\over 2(p-q) C(q,p)^2 m_{2}(\mu_{0,p})}\right)^{1\over 2p+1}
\ .\eqno(5.5)$$
The factor method only requires that $\th(f,t)$ have
$p+2$ continuous derivatives, while the scaling ansatz 
initialization requires $p+4$ derivatives.
The disadvantage of the factor method is that the pilot estimator
requires repeated evaluations of Eq. 5.4 to  determine the minimum,
and therefore is more computationally intensive than the
scalelength initialization.

A third possibility to initialize the multi-stage estimator is to 
fit the data to a low order parametric model. If the evolutionary spectrum
and its derivatives are well described by the parametric model, 
this initialization can yield good, stable results. The 
evolving autoregressive models are of particular interest. 
Several semiparametric (one dimensional) models for fitting the evolutionary
spectrum are  described in Sec. VI.

\np

{\bf VI. Discussion}

The windowed Fourier transform gives estimates of $S(f,t)$ on a grid in
time-frequency space. Smoothing the gridpoint values result in a smooth 
estimate of the evolutionary spectrum. 
Our work generalizes Priestley's original calculation (1966) of the expected 
loss in kernel estimation of evolutionary spectra in several ways. 
First, we  smooth in both time and frequency while Priestley smoothes only in 
time. Second, we explicitly evaluate the optimal kernel halfwidths and
filter bandwidths. Third, we give data-based methods which estimate
the optimal kernel halfwidths with the optimal relative convergence rates.
Finally, we include two effects which do not occur in standard kernel 
estimation problems. 
We include the bias of the gridpoint spectral estimates as a
second order bias correction. Second, as the gridpoint overlap parameters,
$p_T$ and $p_F$, decrease, the correlation in the gridpoint estimates
is included via $\rharg$.

In this article, we have assumed that the signal has no coherent part:
${\rm E}[x_j]= 0$. If unknown coherent components are present,
the resulting model satisfies an evolutionary Cramer's representation:
$$x_t = \sum_{\ell =1}^L A_{ \ell }(t)
\cos({\omega_{\ell } t +\phi_{\ell}} ) + 
\int_{-1/2}^{1/2} A(f,t)e^{2\pi ift}dZ(f)
\ \ , \eqno (6.1)$$
The coherent components, 
$ A_{ \ell }(t) \cos({\omega_{\ell } t +\phi_{\ell}} )$,
may be iteratively estimated aand removed using a kernel smoother 
(Riedel (1992b)).
The error in removing the coherent
component should be included in the the expected loss criteria
for the evolutionary spectrum. However, these additional terms are so
complicated that the residual error from the coherent component removal
is neglected in practice.   
 
The other popular method for smoothing noisy data is smoothing splines.
The simplest smoothing spline approach
is to use crossproduct splines with a knot at every gridpoint,
and with imposed symmetry at $f= 0$.
Convergence results for crossproduct splines are given in
Utreras (1988) and Cox (1984). 
In this formulation, the penalty function is applied uniformly, and 
often tends to penalize the mean spectrum too much and the spectral variation
too little. To remedy this situation, Gu and Wahba (1991) propose a 
nonparametric analysis of variation. In complex demodulation, this 
corresponds to the model: $\th(f,t)= \th_T(t) +\th_F(f) + \tilde{\th}(f,t)$, 
where $\th_T(t)$ and $\th_F(f)$ have smaller penalty functions.
A similar, but different expansion is given in Riedel \& Imre (1993), which
corresponds to the model: $\th(f,t)= \sum_{\ell=1}^Lg_{\ell}(t)\th_{\ell}(f)$,
where the $\{ g_{\ell}(t) \}$ are a small number of known basis functions
such as low order polynomials or complex exponentials. This representation
is parametric in time and nonparametric in frequency.

A third alternative is to perform a singular value decomposition on the
matrix, whose entries are the point estimates of $\th$ evaluated on
the gridpoint locations: $\Theta_{k,j} \equiv \thh(f_k,t_j)$.
The low order left and right eigenvecto
rs are then smoothed to create
the representation: $\th(f,t)= \sum_{\ell=1}^L \th_{T,\ell}(t)\th_{F,\ell}(f)$.
$\th_{T,\ell}(t)$ and $\th_{F,\ell}(f)$ are both nonparametric.

When the spline model is chosen, the strength of the smoothness penalty 
function needs to be selected. The data-based parameter selection method 
which is traditionally used is generalized cross-validation (Wahba (1990)).
Instead, we recommend that the smoothness parameter be chosen to 
minimize the expected loss. Expected loss parameter selection
for smoothing splines with an arbitrary covariance are discussed
in detail in Riedel \& Imre (1993).

The uncertainty principle implies that the time resolution,  $\delta t_R$,
and the frequency resolution, 
$\delta f_R$, satisfy $\delta t_R \delta f_R >{1\over 4 \pi}$.
When the time grid spacing, $\delta t$,
and the frequency grid spacing, $\delta f$, are chosen with 
$\delta t \delta f <{1\over 4 \pi}$, 
the point spectral estimates at the grid locations
are significantly correlated and the effective time-frequency resolution 
continues to satisfy $\delta t_R \delta f_R >{1\over 4 \pi}$. 
Lower values
of the grid spacing parameter, $\delta t \delta f$, are useful, because
the additional estimates of the spectrum contain some  small amount
of information about the spectrum at $(f,t)$. This additional information
does not result in estimates with resolution which exceeds the limit
$\delta t_R \delta f_R ={1\over 4 \pi}$.
 
In complex demodulation, we fix the time grid spacing, $\delta t$,
and the frequency grid spacing, $\delta f$, and then evaluate the point
spectral measurements on an equally spaced grid. By using a single
value of $(\delta t, \delta f)$, we do not use the information 
from other possible
combinations of $\delta t$ and $\delta f$. As $\delta t$ and $\delta f$
are varied, we receive estimates with high frequency resolution and
low time resolution as well as estimates with high time resolution and
low frequency resolution. An improvement on the fixed
$\delta t$ and $\delta f$ approach would be to combine the estimates
from various combinations of $(\delta t, \delta f)$ in a multiresolution
estimate of the evolutionary spectrum. The Wavelet transform
(Daubechies (1990)) yields high time resolution for low frequencies
and poorer time resolution for high frequencies. 
In a future publication, we
plan to examine the wavelet alternative to windowed Fourier transform
estimation of evolutionary spectra.


\noindent
{\bf Acknowledgements}

We thank C. Hurvich, A. Sidorenko, and D.J. Thomson
for useful discussions.
This work was funded by the U.S. Department of Energy.

\np
{\it Bibliography}{}
\begin{enumerate}

\item{Asch, M., Kohler, W., Papanicolaou, G. et al. (1991).
{Frequency content of randomly scattered signals.}
{\it S.I.A.M. Review} {\bf 33}, 519-625.}

\item{
Boashash, B. (1990),
Time-frequency signal analysis.
In {\it Advances in spectrum analysis},
S. Haykin, ed.  Prentice-Hall, New York.}  



\item{
Cohen, L. (1989).
Time-frequency distributions - a review.
{\it  Proc. I.E.E.E.} {\bf 77}, 941-981.}

\item{Cox, D.D. (1984). 
{Multivariate smoothing spline functions.}
{\it SIAM J. Numer. Anal.} {\bf 21}. 789-813. }



\item{Daubechies, I. (1990).
{The wavelet transform, time-frequency localization and signal analysis.}
{\it  I.E.E.E. Trans. on Information Th.} {\bf 36}, 961-1005.}  



\item 
{Grenander, U. and Rosenblatt, M. (1957). 
{\it Statistical analysis of stationary time series.} 
{Wiley}, New York. }

\item{Gu, C. and Wahba, G. (1991).
Smoothing spline ANOVA with componentwise Bayesian ``confidence
intervals''.
Technical Report 881, Department of Statistics, University of
  Wisconsin, Madison.}



\item{ Hardle, W. (1990). {\it Applied nonparametric regression.}
{Cambridge University Press}, Cambridge, New York. }

\item{ Hardle, W., Hall, P., and Marron, S. (1988). 
{How far are automatically chosen smoothing parameters from their optimum?}
{\it J. Amer. Stat. Assoc.} {\bf 83}, 86-95. }



\item{Hlawatsch, F. and Boudreaux-Bartels G.F. (1992).
Linear and quadratic time-frequency representations.
{\it  I.E.E.E. Signal Processing Mag.} {\bf 9}, 21-67.}  



\item 
{Loynes, R.M. (1968). 
{On the concept of spectrum for nonstationary processes.} 
{\it J. Roy. Stat. Soc. Ser. B} {\bf 30}, 1-30.}

\item{Mallows, C.L. (1967).
{Linear processes are nearly Gaussian.}
{\it J. Applied Probability} {\bf 4}, 313-329.}  

\item 
{Melard, G. \& Herteleer-de Schutter, A. (1989). 
{Contributions to evolutionary spectral theory.} 
{\it J. Roy. Stat. Soc. Ser. B} {\bf 10}, 41-63.}


\item{ Mueller, H.G.
(1980). {\it Nonparametric regression analysis of longitudinal data.}
Springer Verlag,  Berlin, New York.}


\item{ Mueller, H.G. and Stadtmueller, U. (1987). 
{Variable bandwidth kernel estimators of regression curves.}
{\it Annals of Statistics} {\bf 15}, 182-201. }

\item{ Papanicolaou, G., Postel, M., Sheng, P.  et al. (1990).
{Frequency content of randomly scattered signals. Part II: Inversion.}
{\it Wave Motion} {\bf 12}, 527-549.}




\item 
{Priestley, M.B. (1965). 
{ Evolutionary spectra and nonstationary processes.} 
{\it J. Roy. Stat. Soc. Ser. B} {\bf 27}, 204-237.} 
\item 
{Priestley, M.B. (1966). 
{ Design relations for  nonstationary processes.} 
{\it J. Roy. Stat. Soc. Ser. B} {\bf 28}, 228-240.} 


\item{Riedel, K.S.
(1992a). 
{Time-frequency tradeoff in data-based parameter determination 
for complex demodulation.}
{\it Proc. I.E.E.E.-S.P. Int. Symp. on Time-Freq. \& Time-Scale Anal.}}

\item{Riedel, K.S. 
(1992b). 
{Time-frequency tradeoff in data-based parameter determination 
for complex demodulation.}
Submitted for publication.} 

\item{Riedel, K.S. and Imre, K. (1993). 
{Smoothing spline growth curves with covariates.}
{\it Comm. in Statistics}, {\bf 22 }, No. 6.  }

\item{Riedel, K.S. and Sidorenko, A. (1993). 
{Optimal edge kernels.} 
Submitted for publication.} 

\item
{Slepian, D. (1978).
{Prolate spheroidal wave functions, Fourier analysis, and
uncertainty-V: the discrete case.}
{\it Bell System Tech. J.} {\bf 57}, 1371-1429.}

\item
{Stone, C.J. (1982).
{Optimal global rates of convergence for nonparametric regression.}
{\it  Annals of Stat.} {\bf 10}, 1040-1053.}


\item
{Thomson, D.J. (1982).
{Spectrum estimation and harmonic analysis.}
{\it Proc. I.E.E.E.} {\bf 70}, 1055-1096.}

\item
{Thomson, D.J. (1990).
{Quadratic inverse spectrum estimates: applications to paleoclimatology.}
{\it Phil. Trans. R. Soc. Lond. A} {\bf 332}, 539-597.}

\item{Thomson, D.J. and Chave, A.D. (1990).
{Jackknife error estimates for spectra, coherences and transfer functions} 
in {\it Advances in spectrum analysis},
(S. Haykin ed.) Ch. 2, {p. 70}, Prentice-Hall, New York.}

\item{Utreras, F. (1988). 
{On generalized cross-validation for 
multivariate smoothing spline functions.}
{\it SIAM J. Sci. Stat. Comp.} {\bf 8}, 630-643. }

\item
{Wahba, G. (1980). {Automatic smoothing of the log periodogram.}
{\it J. Amer. Stat. Assoc.} {\bf 75}, 122-132. 
}

\item{
Wahba, G. (1990).
{\it Spline Models for Observational Data},
S.I.A.M., Philadelphia.}


\item 
{Zurbenko, I.G. (1991). 
{Spectral analysis of nonstationary time series.} 
{\it Int. Stat. Rev.} {\bf 59}, 163-173.} 

\end{enumerate}


\np
\noindent
{\bf Appendix A: Coherence and phase  estimates}

We analyze the cross-coherence and phase 
two separate time series, $\{ x_j^1 \}$ and $\{ x_j^2 \}$.
The tapered Fourier transforms, 
$y_1^{(k)} (f,t)$ and $y_2^{(k)} (f,t)$, are computed from Eq. 2.3. 
The point estimates of the the covariance are computed in the usual
way.
We decompose the cross-coherence into a magnitude and phase:
$$
\hat{C}_{12}(f,t) exp(2\pi i \phih_{\mu}(f,t) = 
{\hat{S}_{12} (f) \over \sqrt{\hat{S}_{11} (f) \hat{S}_{22} (f)}}
\eqno (A1)
$$
The cross-coherence estimate is stabilized with the
inverse hyperbolic tangent transformation
(Thomson and Chave (1990)):
$$
Q(f,t) \equiv \sqrt{4K-2} \ {\rm tanh}^{-1} (|C_{12}(f,t) |) \ , \eqno (A 2)
$$
where $K$ is the number of tapers of different tapers. 
In our single taper point estimates, $K = 1$.
The ${\rm tanh}^{-1}$ transform accelerates the convergence of 
the residual errors to a Gaussian. 
When $\{ y^1 \}$ and $\{ y^2 \}$ are jointly Gaussian, the bias and
variance of $Q$ are explicitly known: 
$Q$ has a small bias,
$E \hat{Q} = Q + {1 \over \sqrt{4K -2} }$, and unit variance.
After smoothing, $Q(f,t)$ and its confidence interval are then
transformed back to the coherence.

To compute the phase estimates, we smooth the real and imaginary
parts of $\gamh(f,t)\equiv exp(2\pi i \phih(f,t))$. The smoothed
estimate of $\gam(f,t)$ has a modulus less than one. Therefore,
our smoothed estimate of the phase is defined by 
$exp(2\pi i \phih_{\mu}(f,t) \equiv \gamh_{\mu}(f,t)/|\gamh_{\mu}(f,t)|$.

\end{document}